\documentclass[twocolumn,showpacs,preprintnumbers,amsmath,amssymb]{revtex4}
\usepackage{graphicx}
\usepackage{dcolumn}
\usepackage{bm}
\usepackage{gensymb}

\begin{document}


\title{Multistage Zeeman deceleration of NH $X\,^3\Sigma^-$ radicals}
\author{Vikram Plomp}
\author{Zhi Gao}
\author{Theo Cremers}
\author{Sebastiaan Y.~T.~van de Meerakker}
\affiliation{Radboud University, Institute for molecules and materials, Heijendaalseweg 135, 6525 AJ Nijmegen, the Netherlands}

\begin{abstract}
We report on the Zeeman deceleration of ground-state NH radicals, using a decelerator that consists of 100 pulsed solenoids and 100 permanent hexapoles. Packets of state-selected NH ($X\,^3\Sigma^-$, N=0, J=1) radicals are produced with final velocities ranging between 510 m/s and 150 m/s. The velocity distributions of the packets of NH exiting the Zeeman decelerator are probed using velocity map imaging detection. We present a new 1+2' resonance-enhanced multiphoton ionization scheme for NH, that allows for velocity map imaging detection under ion recoil-free conditions. The packets of Zeeman-decelerated NH radicals, in combination with the new detection scheme, offer interesting prospects for the use of this important radical in high-resolution crossed-beam scattering experiments.
\end{abstract}

 \maketitle

\section{Introduction}
In recent years, there has been a growing interest to study collisions between atoms and molecules under well-controlled conditions, and at low collision energies. This interest stems from the exotic and intriguing quantum phenomena that can be observed, such as the occurrence of quantum diffraction oscillations \cite{Zastrow:NatChem6:216,Vogels:PRL113:263202,Onvlee:NatChem9:226}, product-pair correlations in bimolecular collisions \cite{Gao:NatChem10:469} or scattering resonances \cite{Chandler:JCP132:110901,Chefdeville:Science341:06092013,Henson:Science338:234,Vogels:SCIENCE350:787}. The experimental observation of these phenomena provide extremely sensitive tests for quantum scattering calculations \cite{Bell:MolPhys107:99,Vogels:NatChem10:435}, and paves the way for future research directions that aim to control collisions or chemical reactions using externally applied electric or magnetic fields \cite{Krems:PCCP10:4079}.

Nowadays, various methods are available to study molecular collisions at temperatures below a few Kelvin. Molecules can be confined in traps where collisions may be studied between the molecules \cite{Stuhl:NATURE492:396,Wu:Science358:645,Hummon:PRL106:053201}, or with a co-trapped collision partner \cite{Akerman:PRL119:073204}. The crossed-beam technique, historically the workhorse to study molecular collisions with the highest possible level of precision \cite{Lee:AC26:939}, can also be used to reach the required low energies, either by allowing the beams to cross at small or even zero intersection angle \cite{Chefdeville:Science341:06092013,Henson:Science338:234,Osterwalder:EPJ-TI2:10}, by slowing the molecules down using Stark or Zeeman decelerators \cite{Meerakker:CR112:4828,Hogan:PCCP13:18705}, or by using a combination of both.

The NH radical has been a species of primary interest for these experiments, ever since the field of cold molecules developed since the late 90's. NH in the metastable $a\,^1\Delta$ state is very amenable to the Stark deceleration technique, whereas the 2 $\mu_B$ magnetic moment in the $X\,^3\Sigma^-$ electronic ground state makes the molecule a prime candidate for magnetic deceleration and trapping experiments. The electronic energy level structure of NH in principle allows for the re-loading of packets of ground-state NH ($X\,^3\Sigma^-$) molecules in a magnetic trap after Stark deceleration in the metastable $a\,^1\Delta$ state, thereby increasing the density of trapped molecules \cite{Meerakker:PRA64:041401}. Stark deceleration \cite{Meerakker:JPB39:S1077} and electrostatic trapping of NH ($a\,^1\Delta$) \cite{Hoekstra:PRA76:063408}, as well as the subsequent transfer of multiple packets into a magnetic trap was experimentally achieved, demonstrating the feasibility of the approach \cite{Riedel:EPJD65:161}. Trapped samples of NH ($X\,^3\Sigma^-$) have also been produced via the buffer gas cooling technique \cite{Campbell:PRL98:213001}, and used to measure the radiative lifetime of vibrationally excited NH \cite{Campbell:PRL100:083003}.  The electric dipole allowed $A\,^3\Pi \leftarrow X\,^3\Sigma^- $ transition has a remarkably high Franck-Condon factor \cite{Meerakker:PRA64:041401}, which offers interesting prospects for direct laser-cooling. Collisions involving cold samples of NH have also been the subject of a number of high-level theoretical investigations \cite{Janssen:JCP131:224314,Janssen:JCP134:124309,Janssen:EPJD65:177,Janssen:PRA83:022713,Janssen:PRL110:063201}. NH is expected to have interesting collision properties with Rb atoms due to near-resonant electronic energy levels \cite{Haxton:PRA80:022708}. Recent work also suggests that NH may have favorable collision properties to be used in sympathetic cooling approaches \cite{Wallis:EPJD65:151,Zuchowski:PCCP13:3669}. Furthermore, NH is an important species in astrochemistry, and collisions between NH and He atoms or H$_2$ molecules have been investigated theoretically, predicting the occurrence of pronounced scattering resonances at temperatures below 10 K \cite{Ramachandran:JCP148:084311}\cite{Bouhafs:JCP143:184311}. Last but not least, a large body of work exists on state-to-state crossed-beam experiments involving NH collisions at higher energies \cite{Dagdigian:JCP90:6110,Rinnenthal:JCP113:6210,Rinnenthal:JCP116:9776}.

Despite its relevance and amenability to the Zeeman deceleration technique, the deceleration of ground-state NH ($X\,^3\Sigma^-$) radicals using a Zeeman decelerator has not yet been reported. This is in part due to the fact that a relatively long Zeeman decelerator is required to significantly reduce the mean velocity of a beam of NH, and in part due to the relatively difficult production and detection techniques for NH. Here, we present the deceleration of ground-state NH ($X\,^3\Sigma^-$) molecules using a 3-meter-long Zeeman decelerator that consists of 100 pulsed solenoids and 100 permanent hexapoles. Packets of state-selected NH in the rovibrational $v=0, N=0, J=1$  ground state are produced with a continuously variable mean velocity in the 510-150 m/s range. The packets of NH radicals exiting the decelerator are characterized using Velocity Map Imaging (VMI) detection, directly probing both the longitudinal and transverse velocity spreads of the molecular distributions. To record these velocity spreads with a high resolution, we present a new 1+2' Resonance-enhanced Multi Photon Ionization scheme (REMPI) with which NH radicals are state-selectively ionized under ion recoil-free conditions.

\section{Experiment}
\begin{figure}[!htb]
\centering
\resizebox{1.0\linewidth}{!}
{\includegraphics{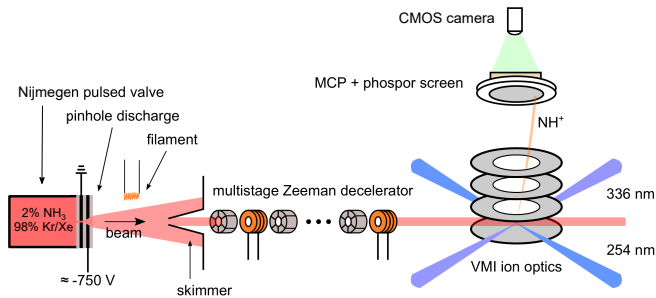}}
\caption{(color online). Schematic depiction of the experimental setup. NH molecules in the $X^{3}\Sigma^-$ electronic ground state are created by an electric discharge of NH$_3$
    seeded in either Kr or Xe. The beam is collimated by a skimmer, and then passed through a Zeeman decelerator consisting of an alternating array of 100 pulsed solenoids and 100 permanent magnets. The NH radicals exiting the Zeeman decelerator are guided into the detection region by a number of additional hexapoles (not shown), and detected using velocity map imaging. State-selective and ion recoil-free ionization is achieved using a two-color 1+2' REMPI scheme employing two tunable dye lasers.}
\label{fig:setup}
\end{figure}

The experiments are performed in a newly constructed molecular beam apparatus containing a 3-meter-long Zeeman decelerator that is schematically shown in Figure \ref{fig:setup}. This apparatus has been used recently to decelerate beams of O atoms and O$_2$ molecules \cite{Cremers:PRA98:033406}. Here, we restrict ourselves to a description of the experimental details that are specific to the production, deceleration and detection of the NH radical.

A molecular beam of NH ($X\,^3\Sigma^-$) molecules with a forward velocity centered around 550 or 430 m/s is formed by an electric discharge of 2\% NH$_{3}$ seeded in krypton or xenon, respectively, using a Nijmegen Pulsed Valve with discharge assembly \cite{Ploenes:RSI87:053305}. This valve is operated at a repetition rate of 10~Hz. After the supersonic expansion, most NH radicals in the $X\,^3\Sigma^-$ electronic ground state reside in the $v=0, N=0, J=1$ rovibrational ground state. This state has a magnetic moment of 2 $\mu_B$, and splits into a $M_J=1$, $M_J=0$ and $M_J=-1$ component in the presence of a magnetic field. Only NH radicals in the low-field seeking $M_J=1$ component are selected by the Zeeman decelerator in the experiments, although molecules in the $M_J=0$ component are insensitive to magnetic fields and pass through the Zeeman decelerator in free flight. Their density, however, is heavily reduced during this flight, such that their contribution is negligible under most conditions.

Approximately 90~mm downstream from the nozzle orifice, the molecular beam passes a 3~mm diameter skimmer, and enters the Zeeman decelerator. The Zeeman decelerator consists of an alternating array of 100 solenoids and 100 hexapoles. A detailed description of the mechanical and electronical implementation is given elsewhere \cite{Cremers:RSI}. Briefly, the decelerator consists of 5 modules containing 20 solenoids and 19 hexapoles each, that are connected to each other using a hexapole positioned at the interface between the modules. The solenoids are made of copper capillary through which currents up to 5~kA are pulsed. Cooling liquid is passed through the solenoids. Each solenoid is connected to an individual printed circuit board to provide the current pulses using FET-based electronics components. The hexapoles consist of six commercially available arc-shaped permanent magnets.

The NH ($X\,^3\Sigma^-$) radicals that exit the Zeeman decelerator are state-selectively detected using REMPI. Depending on the experiment, different REMPI schemes are used. To record time-of-flight (TOF) profiles of the radicals exiting the decelerator, we use a 2+1 REMPI scheme. In this scheme, NH is first resonantly excited to the $D\,^3\Pi$ state using photons at a wavelength around 224 nm, before they are ionized by absorption of a third photon of the same wavelength \cite{Clement:JCP97:7064}.

Although rather efficient, this 2+1 scheme via the $D$ state imparts a recoil velocity of about 38~m/s to the ions, significantly blurring the recorded images when VMI is used. This is inconsequential for the measurement of TOF profiles, but it is detrimental in, for instance, measurements of scattering images in a crossed-beam experiment. We therefore developed an alternative ion recoil-free 1+2' REMPI scheme. In this scheme, we first resonantly excite NH to the $A\,^3\Pi$ state via the strong $A \leftarrow X$ transition, using photons at a wavelength around 336 nm. The NH radicals are then ionized through the absorption of 2 photons at a wavelength around 254 nm that are provided by a second tunable dye laser.

In this 1+2' REMPI scheme, the wavelength of the ionization laser can be tuned to the ionization threshold, offering a direct route to recoil-free detection of NH. However, the ionization step is found to be rather inefficient. The electron configuration of NH ($A\,^3\Pi$) is given by $1\sigma^2 2\sigma^2 3\sigma 1\pi^3$, whereas the configuration for the $X\,^2\Pi$ ground-state NH$^+$ ion is given by $1\sigma^2 2\sigma^2 3\sigma^2 1\pi$, i.e., two electron transitions are involved in the ionization step \cite{Amero:IJQC99:353}. We found that the ionization efficiency of this 1+2' REMPI scheme is enhanced by more than an order of magnitude if the frequency of the ionization laser is blue-detuned by about 600 cm$^{-1}$ with respect to the ionization threshold. The origin of this enhanced efficiency is at present unclear, and more spectroscopic work is needed to elucidate the exact ionization pathway for this peak. We speculate, however, that it could be related to resonant excitation to a Rydberg state that auto-ionizes either to the $X\,^2\Pi$ ground-state of the NH$^+$ ion, or to the close-lying $a\,^4\Sigma^-$ first excited state of NH$^{+}$. The latter has electron configuration $1\sigma^2 2\sigma^2 3\sigma 1\pi^2$ \cite{Amero:IJQC99:353}, such that during the ionization process only a $\pi$ electron has to be excited. We have observed that this more efficient ionization pathway is accompanied by a small ion recoil velocity of a few m/s. When using the 1+2' REMPI scheme, we therefore have a trade-off between ionization efficiency and ion recoil velocity. By simply changing the wavelength of the ionization laser, signal intensities can be enhanced at the expense of a few m/s velocity blurring.

For the 2+1 REMPI scheme via the $D$ state, the 224~nm light is generated by frequency tripling the output of a dye laser (LiopTec), pumped using the second harmonic of an injection seeded Nd:YAG laser (Continuum Surelite EX). Typically, a 5~mm diameter laser beam with a pulse energy of 2 mJ in a 5~ns pulse is used, that is focused in the interaction region using a spherical lens with 500 mm focal length. For the two-color 1+2' REMPI scheme, the (unseeded) Nd:YAG laser is used to pump two dye lasers simultaneously. The second harmonic of the Nd:YAG laser is used to pump a Spectra Physics PDL2 dye laser, which is frequency doubled to produce radiation at 336 nm. The $A \leftarrow X$ transition in NH is easily saturated, and only low laser powers ($< 1$ mJ) are needed. The second color is produced by pumping a LiopTec dye laser by the third harmonic of the Nd:YAG laser. The output of this laser is frequency doubled to generate laser radiation at wavelengths around 254 nm. Typically, a 5~mm diameter laser beam with a pulse energy of 1.5 mJ in a 5~ns pulse is used, that is focused in the interaction region using a spherical lens with 400 mm focal length. A delay line is installed in the beam path for the second color to make sure that both lasers intercept the interaction region within the 449~ns lifetime of the $A \,^3\Pi$ state \cite{SONG:CTC1093:81}.

After the REMPI process, the NH ions are detected with a VMI detector, of which the ion optics consist of a repeller, two extractors, and a grounded plate. A repeller voltage of 2000 V is used to accelerate the ions towards a microchannel plate (MCP) detector. Impact positions of impinging ions are recorded by a phosphor screen in combination with a CMOS camera and home-written acquisition and analysis software. For the measurement of TOF profiles, the VMI detector is operated out of velocity focus, and the integral signal recorded by the camera is used. For the measurement of velocity distributions, the VMI detector is operated in velocity focus, such that 2D images that reflect the velocity distributions of the molecular packets are generated.

\section{Results and discussion}
\subsection{Time-of-flight measurements}
The Zeeman decelerator is used in two modes of operation. In the so-called hybrid mode \cite{Cremers:PRA95:043415}, each solenoid is pulsed twice to guide a packet of NH radicals through the decelerator at constant speed. This mode of operation is very similar to operation of a Stark decelerator in guiding mode \cite{Bethlem:PRL84:5744}. In the so-called deceleration mode, only a single current pulse is passed through each solenoid. Depending on the exact timings, governed by the so-called phase angle $\phi_0$ \cite{Meerakker:CR112:4828,Cremers:PRA95:043415}, a near constant amount of kinetic energy is removed from the packet of NH per solenoid. The packet therefore exits the decelerator at a lower longitudinal velocity, which is controlled by the pulse-sequence of the solenoids only.

\begin{figure}[!htb]
\centering
\resizebox{1.0\linewidth}{!}
{\includegraphics{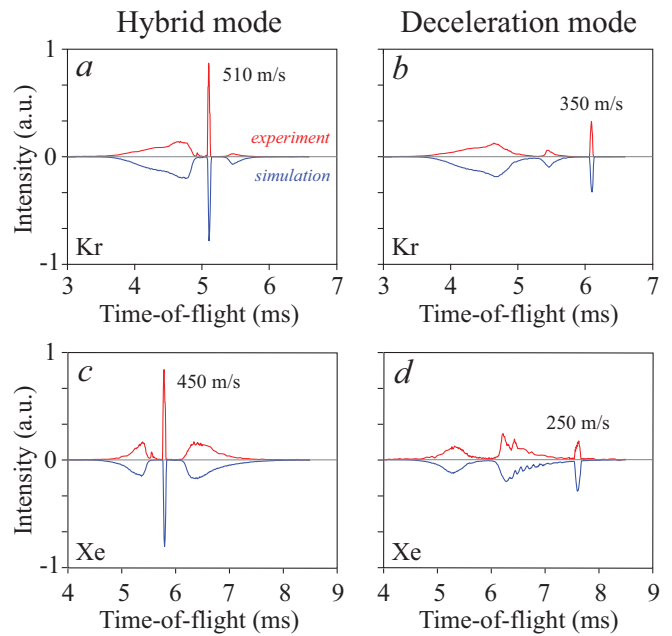}}
\caption{(color online). Time-of-flight profiles for NH radicals ($X\,^3\Sigma^-$) exiting the Zeeman decelerator. Either krypton (upper panels) or xenon (lower panels) is used as a carrier gas to produce the initial beam of NH. The decelerator is used in hybrid mode (panels \emph{a} and \emph{c}) or in deceleration mode (panels \emph{b} and \emph{d}). In each panel, the experimentally observed profile (red curves) is shown above the simulated profile (blue curves).}
\label{fig:fullTOF}
\end{figure}

Typical TOF profiles of NH ($X\,^3\Sigma^-, N=0, J=1$) radicals that are observed when the Zeeman decelerator is operated in hybrid mode and deceleration mode are shown in Figure \ref{fig:fullTOF}. Krypton (panels \emph{a} and \emph{b}) or xenon (panels \emph{c} and \emph{d}) is used as a carrier gas in these experiments. In both modes of operation, the decelerator is programmed to select a packet of NH with a mean speed of 510~m/s (Kr) or 450 m/s (Xe). In hybrid mode, the selected packet is transported through the decelerator, while keeping the packet together in both the longitudinal and transverse directions. It appears in the TOF profile as a narrow and intense peak that is well separated from the remainder of the gas pulse. In deceleration mode, the decelerator is programmed to reduce the mean speed of the selected part to 350~m/s (Kr) or 250 m/s (Xe). The decelerated packet of NH is detected much later, and is seen to be split off from the part of the beam that is not decelerated.

All TOF profiles show very good agreement with the TOF profiles that are obtained from three dimensional trajectory simulations of the experiment, that are shown underneath the experimental profiles. In these simulations, the magnetic field distributions generated by the solenoids and hexapoles are taken into account, as well as the temporal current profiles applied to the solenoids. The initial beam of NH is simulated using Gaussian spatial and velocity spreads. In the simulations, the NH radicals are assumed to exclusively reside in the $M_J=1$ component of the $N=0, J=1$ state. All main features of the experimentally observed TOF profiles are well reproduced by the simulations, indicating that the transport of the molecules through the decelerator is accurately described by the simulations and well understood.

\begin{figure}[!htb]
\centering
\resizebox{1.0\linewidth}{!}
{\includegraphics{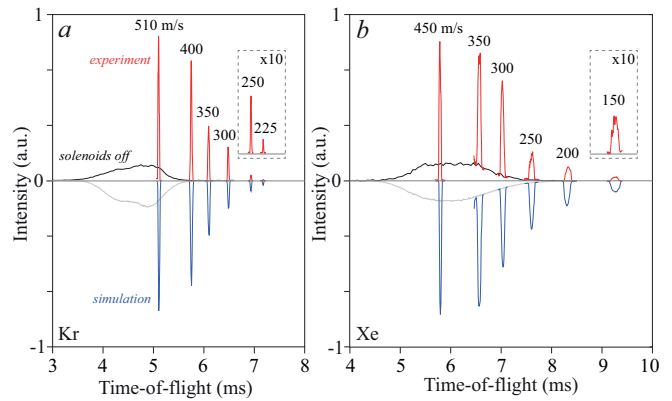}}
\caption{(color online). Selected parts of time-of-flight profiles for NH radicals ($X\,^3\Sigma^-$) exiting the Zeeman decelerator when the decelerator is programmed to produce packets of NH with different final velocities. Either krypton (\emph{a}) or xenon (\emph{b}) is used as a carrier gas to produce the initial beam of NH. In each panel, the experimentally observed profiles (red curves) are shown above the simulated profiles (blue curves). The experimental and simulated profiles that are observed when all solenoids are turned off are shown in black and grey, respectively.}
\label{fig:peakTOF}
\end{figure}

The final velocity of the decelerated packet can be tuned by changing the phase angle with which the decelerator is operated, or by selecting a different initial velocity of the beam pulse. In figure \ref{fig:peakTOF} a series of TOF profiles is shown that is observed when the Zeeman decelerator is programmed to  produce packets of NH with a final velocity ranging between 510 m/s and 150 m/s. The profiles in panel \emph{a} and \emph{b} are observed when molecular beams of NH are produced using Kr and Xe as a carrier gas, again selecting initial speeds of 510 m/s and 450 m/s, respectively. Only the parts of the TOF profiles containing the selected packets are shown. For reference, the peaks that are observed when the decelerator is operated in hybrid mode are shown again. In addition, the profiles are shown that are observed when the solenoids are not activated, i.e. when the beam is only transversally focused by the hexapoles. The profiles that result from the numerical trajectory calculations are shown underneath the experimental profiles. Again, very good agreement between experiment and simulation is obtained, both with respect to the arrival time distribution of the packet, and with respect to the relative intensities of the peaks.

It is notoriously difficult to estimate particle densities from the observed signal levels. We therefore refrain from such an estimate; but note that typically a few hundred ions per shot are detected when the 2+1 REMPI scheme is used, the Zeeman decelerator is operated in hybrid mode, and Kr is used as a seed gas. These signal levels appear relatively large in comparison to the Zeeman deceleration of O atoms and O$_2$ molecules reported earlier \cite{Cremers:PRA98:033406}.

\subsection{Measurements of velocity distributions}
The velocity-controlled packets of NH that exit the Zeeman decelerator are analyzed further by recording their velocity distributions using the VMI detector. For different settings of the decelerator, the NH packet is probed at the time where the time-of-flight profile has maximal intensity. We refer to these as beamspot measurements. For these measurements, we use the 1+2' REMPI scheme, where we select the frequency of the second color to ionize NH just above threshold. Furthermore, in order to avoid potential blurring due to excessive signal levels and Coulomb repulsion, we work with attenuated laser powers such that a maximum of one ion per shot is recorded.

\begin{figure}[!htb]
\centering
\resizebox{1.0\linewidth}{!}
{\includegraphics{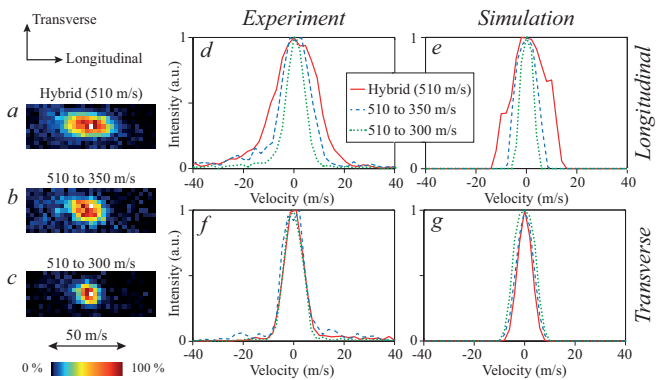}}
\caption{(color online). Measured velocity mapped ion images of NH radicals ($X\,^3\Sigma^-$) that exit the Zeeman decelerator if the decelerator is operated in hybrid mode selecting a packet with a mean velocity of 510 m/s (\emph{a}), or in deceleration mode generating a packet with a final velocity of 350 m/s (\emph{b}) or 300 m/s (\emph{c}). Experimental longitudinal (\emph{d}) and transverse (\emph{f}) velocity distribution, together with the corresponding simulated distributions (\emph{e, g}).}
\label{fig:beamspots}
\end{figure}

In Figure \ref{fig:beamspots} three such beamspots are shown that are recorded when Kr is used as a seed gas to produce the initial beam of NH. The beamspots are presented such that the longitudinal and transverse directions are oriented horizontally and vertically, respectively. In panel \emph{a}, the Zeeman decelerator is operated in hybrid mode, selecting a packet of NH with a mean velocity of 510~m/s. Panels \emph{b} and \emph{c} show the beamspots that are recorded when the decelerator is operated in deceleration mode using an effective phase angle of about 0$\degree${} and 29$\degree${}, producing a packet of NH with a final velocity of 350 m/s and 300 m/s, respectively. It is observed that the longitudinal velocity distribution is widest when the Zeeman decelerator is operated in hybrid mode, and becomes increasingly narrow when the NH packet is decelerated to lower final velocities. The transverse spread, however, is observed to be almost identical for all modes of operation.

The shapes of the recorded beamspots directly reflect the velocity distributions of the NH packets, which can be analyzed quantitatively. For this, we carefully calibrated the VMI detector using a procedure originally developed for Stark decelerators, which entails the measurements of a series of beamspots as a function of the final velocity of the decelerated packet \cite{Onvlee:PCCP16:15768}. For the VMI detector used here, we find that one pixel in the images corresponds to a velocity of 2.3~m/s.

The resulting longitudinal and transverse velocity distributions of the three beamspots are shown in panels \emph{d} and \emph{f}, respectively. The width (full width at half maximum) of the longitudinal distributions ranges between 20~m/s and 8~m/s, whereas all transverse distributions have a width of about 9~m/s. As a comparison, the corresponding distributions resulting from the numerical trajectory calculations are shown in panels \emph{e} and \emph{g}. A very good agreement between the experimental and simulated profiles is observed. Qualitatively, the decreasing width of the longitudinal distributions and the insensitivity of the observed transverse distributions to the different modes of operation is reproduced well by the simulations. Quantitatively, the measured profiles have a little larger width than the simulated profiles. We mainly attribute this to small blurring effects of the non-perfect velocity mapping conditions of our VMI detector. Furthermore, the simulated magnetic field distributions of hexapoles and solenoids might deviate from the actual ones, and the simulated molecular beam distribution at the entrance of the decelerator may not perfectly describe the actual distribution present in the experiment.

The observed trends in the velocity distributions can be directly understood from the operation principles of the Zeeman decelerator. The longitudinal velocity spread of the selected packet is governed by the longitudinal phase-space acceptance of the decelerator, given by the so-called separatrix \cite{Cremers:PRA95:043415}. This separatrix is largest for hybrid mode, and becomes smaller as the final velocity of the packet is reduced in deceleration mode. The transverse spread is almost exclusively governed by the hexapoles; the solenoids have only a minor effect on the transverse motion. This transverse motion is thus almost independent on the currents passed through the solenoids, resulting in a near-identical transverse distribution in all cases. These observations are consistent with the design concept of our Zeeman decelerator, that entails the decoupling of the longitudinal deceleration and transverse focusing properties \cite{Cremers:PRA95:043415}. The resulting unequal partitioning of the decelerator acceptance between the longitudinal and transverse directions, as demonstrated here experimentally, is ideal in scattering experiments.

\section{conclusions}
We have presented the successful Zeeman deceleration of ground-state NH ($X\,^3\Sigma^-, N=0, J=1$) radicals, using a Zeeman decelerator consisting of 100 solenoids and 100 hexapoles. Packets of NH with well-defined spatial, velocity and temporal spreads are produced with final velocities in the 510-150~m/s range. We presented a new 1+2' resonance-enhanced multiphoton ionization scheme, that allows for velocity map imaging detection of NH under ion recoil-free conditions. We have used this detection scheme to experimentally determine the velocity spreads of the packets of NH exiting the Zeeman decelerator. These measurements confirm that the longitudinal velocity spread is governed by the mode of operation of the decelerator, whereas the transverse spread is independent of the operation mode and almost exclusively determined by the hexapoles. This work is primarily aimed at the use of NH in controlled crossed-beam scattering experiments, but the Zeeman deceleration of NH and the 1+2' REMPI scheme presented here may find applications in experiments that aim for the trapping of NH after Zeeman deceleration as well.

\section{Acknowledgements}
The research leading to these results has received funding from the European Research Council under the European Union's Seventh Framework Programme (FP7/2007-2013/ERC grant agreement nr. 335646 MOLBIL). This work is part of the research program of the Netherlands Organization for Scientific Research (NWO). We thank Simon Chefdeville for help in the development of the NH production and detection techniques. We thank Mike Ashfold for stimulating discussions on REMPI of NH. We thank Niek Janssen, Andr\'e van Roij and Edwin Sweers for expert technical support.

\bibliographystyle{apsrev}
\bibliography{../bib/string,../bib/mp}

\end{document}